\documentclass[twocolumn,superscriptaddress,amsmath,amssymb,aps,floatfix,nofootinbib]{revtex4-2}

\usepackage{overpic}
\usepackage{graphicx}
\usepackage{multirow}
\usepackage{dcolumn}
\usepackage{bm}
\usepackage{float} 
\usepackage{xcolor}
\usepackage{soul}
\usepackage{diagbox}
\usepackage{booktabs}

\begin{document}

\title{Nonequilibrium Phenomenology of Identified Particle Spectra in\\ Heavy-Ion Collisions at LHC Energies}

\author{Oleksandr Vitiuk}
\affiliation{Institute of Theoretical Physics, University of Wroc\l{}aw, Plac Maxa Borna 9, 50-204 Wroc\l{}aw, Poland}
\author{David Blaschke}
\affiliation{Institute of Theoretical Physics, University of Wroc\l{}aw, Plac Maxa Borna 9, 50-204 Wroc\l{}aw, Poland}
\affiliation{Center for Advanced Systems Understanding (CASUS), Untermarkt 20, D-02826 G{\"o}rlitz, Germany}
\affiliation{Helmholtz-Zentrum Dresden-Rossendorf (HZDR), Bautzener Landstrasse 400, D-01328 Dresden, Germany}
\author{Benjamin D{\"o}nigus}%
\affiliation{Institut f{\"u}r Kernphysik, Goethe Universit{\"a}t Frankfurt am Main, Max-von-Laue-Str. 1, 
D-60438 Frankfurt am Main, Germany}%
\author{Gerd R{\"o}pke}
\affiliation{Institute of Physics, University of Rostock, Albert-Einstein Str. 23-24, D-18059 Rostock, Germany}

\date{\today}

\begin{abstract}
{
We employ the Zubarev approach of the non-equilibrium statistical operator to investigate the enhancement of the low-$p_T$ region of pion spectra, introducing an effective pion chemical potential to describe the overpopulation of low-energy pion states. 
We test a corresponding freeze-out approach by analyzing the transverse-momentum spectra of identified particles measured recently with high precision by the ALICE Collaboration in Pb+Pb collisions at CERN LHC. 
A blast-wave model and a blast-wave-based particle generator, coupled to a hadronic transport model, are utilized. Bayesian inference methods are applied to extract the most probable sets of thermodynamic parameters at the chemical freeze-out hypersurface. Both models for the overpopulated pion states, the hadronic transport model and the thermal model with a nonzero pion chemical potential, provide a satisfactory description of the observed pion spectra.
However, both approaches contain approximations which can be improved within a systematic nonequilibrium approach.   We demonstrate that the introduction of a nonequilibrium pion chemical potential offers an efficient alternative to the conventional explanation of the low-$p_T$ enhancement, typically attributed to resonance decays with subsequent thermalization. 
A similar discussion holds also for the kaon spectra.
}
\end{abstract}

\maketitle

\section{Introduction}\label{sec:intro}

Since the early days of ultrarelativistic heavy-ion collisions at CERN-SPS the yields and spectra of identified particles have been the key observables to deduce the thermodynamic properties of the state of matter that has been created in these collisions \cite{Satz:1988pn}.
In order to understand the emergence of hadronic yields it has been expected that a statistical description following Hagedorn \cite{Hagedorn:1965st,Hagedorn:1968zz} could be applicable.
Later it was demonstrated that the ratios of particle yields from the first CERN-SPS experiments could nicely be described within a simple chemical freeze-out model from the statistical equilibrium state of a hadron resonance gas with just two parameters, the freeze-out temperature and freeze-out baryon chemical potential \cite{Braun-Munzinger:1995uec}.
However, as a caveat of such a statistical model of hadron production, the spectra of pions (and kaons) at low transverse momenta could not be described well with simple thermal distribution functions \cite{Strobele:1988,Kataja:1990tp}. 
Compared with these distribution functions, the low-$p_T$ region of observed  pion spectra are overpopulated.

A nonequilibrium distribution function with a finite pion (and kaon) chemical potential could well describe the observed enhancement in the low transverse momentum part of the spectrum as a precursor of Bose-Einstein condensation \cite{Kataja:1990tp}.
Alternatively, the feed-down from resonance decays and the effect of radial flow has been considered \cite{BWModel} and gave at that time a satisfactory description of the spectra which afterwards has been adopted as a standard tool in the theoretical description of particle spectra. 

With the advent of high-precision and high-statistics data of identified particle production from the ALICE experiment at CERN-LHC \cite{ALICE-2013,ALICE:2019hno},
the problem of a precise description of the low-momentum pion spectra came up again since the modern hydrodynamic codes systematically underestimate the pion yield at low transverse momentum \cite{heinz2024bayesian, Devetak2020, Everett2021}. The same problem is present in the blast-wave fits of the experimental data and the simple feed-down from resonance decays is insufficient \cite{ALICE-2013,Mazeliauskas_PRC_2020}. As a possible explanation, the critical enhancement of soft pions by fluctuations at the chiral transition \cite{Grossi:2021gqi} was proposed. While the estimated enhancement factor is in agreement with the
observed values, the direct comparison with the data was never done. Another explanation was given in the framework of the nonequilibrium statistical hadronisation model with quark fugacity factors \cite{Begun_PRC_2014}. However, hadronic rescatterings in the
final stage were neglected.
We suggest that the new high precision {CERN-LHC} data require a more fundamental approach.

In this work, we present a systematic nonequilibrium approach to the pion production within the Zubarev method of the nonequilibrium statistical operator (NSO) \cite{Zubarev} and apply it to a description of the recent experimental data presented by the ALICE Collaboration \cite{ALICE-2013,ALICE:2019hno}.
To this end we will develop and use a generalized blast-wave model and discuss different approximations, such as the introduction of
a nonequilibrium chemical potential, and include also the effects of radial flow, resonance decays and final state interactions.

\section{The method of nonequilibrium statistical operator (NSO)}
\label{sec:zubarev}

To determine the statistical operator $\rho(t)$ for a nonequilibrium process, we have to solve the Liouville-von Neumann equation
\begin{equation}
   \frac{\partial}{\partial t}\rho(t)=\frac{i}{\hbar}[H,\rho(t)]
\end{equation}
with given initial conditions.
These initial conditions are represented by the average values of a set of relevant observables $\{B_i\}$ in the past $t'\le t$ which characterize the state of the system. This information is used to construct the relevant statistical operator $\rho_{\rm rel}( t')$ as the maximum of information entropy
\begin{equation}
    S_{\rm inf}(t')=-\langle \ln[\rho_{\rm rel}(t')]\rangle_{\rm rel}    
\end{equation}
under given constraints, the self-consistency conditions
\begin{equation}
\label{selfc}
   \langle B_i \rangle^{t'} = {\rm Tr}\left\{\rho_{\rm rel}( t') B_i\right\}.
\end{equation}
As well known from equilibrium statistics, these self-consistency conditions are taken into account via Lagrange parameters $\lambda _i(t')$,
and we obtain the generalized Gibbs distribution
\begin{equation}
    \rho_{\rm rel}(t')=\dfrac{e^{-\sum_i \lambda_i(t') B_i}}{{\rm Tr}e^{-\sum_i \lambda_i(t') B_i}}.
\end{equation}
The Lagrange parameters $\lambda _i(t')$ must be eliminated using the self-consistency conditions (\ref{selfc}) which are the nonequilibrium generalizations of the equations of state.

The solution of Eq. (\ref{selfc}) is not trivial since the relevant operators can contain few-particle terms like the interaction in the Hamiltonian. We can apply many-particle techniques. A spectral function can be introduced, perturbation theory with respect to the interaction in the Hamiltonian gives the excitations of the system including the medium effects. For instance, the resonance gas represents a simple approximation. Note that a systematic approach requires the account of in-medium effects and continuum correlations, as needed, for instance to explain the high-precision data {for the yields measured in} the ALICE experiment \cite{Andronic:2018qqt,Donigus:2022haq}.

Below we discuss the density as relevant observable, depending on temperature and chemical potential (Lagrange parameters). We mention three approximations to treat the spectral function: (a) the ideal quantum gas, (b) the resonance-gas approximation where excited states are treated in chemical equilibrium, and (c) the inclusion of in-medium effects such as the quasiparticle shifts, Pauli blocking, continuum correlations as known from the generalized Beth-Uhlenbeck formula \cite{Schmidt:1990oyr} and the description of the Mott effect for clusters \cite{Blaschke:2024jqd}.

The solution of the Liouville-von Neumann equation at given boundary conditions (\ref{selfc}) is \cite{Zubarev}
\begin{equation}
\label{rhoZ}
\rho(t)=\lim_{\epsilon \to 0} \epsilon \int_{-\infty}^{t} \hspace{-3mm}d t' e^{-\epsilon ( t- t')}
e^{-\frac{i}{\hbar} H ( t- t')}\rho_{\rm rel}( t')e^{\frac{i}{\hbar} H ( t- t')}.
\end{equation}
Having $\rho(t)$ to our disposal, we can follow the time evolution of the system. In particular, we can calculate transport coefficients (reaction rates) \cite{Zubarev}. Within a quantum statistical approach, in simplest approximation the free reaction rates are obtained. Improved approximations lead to medium modifications of the transport coefficients, see for example Ref. \cite{Barker:2016hqv}.

A particular problem of the method of NSO is the selection of the set of relevant observables $\{B_i\}$. We have to consider long-living fluctuations
because they need a long time to be built up dynamically and demand higher orders of perturbation theory. 
A minimum set of relevant observables consists of the conserved quantities: energy $H$ as well as the conserved numbers $N_i$.
The Gibbs distribution
\begin{equation}
    \rho_{\rm rel}(t) \propto \exp[-(H-\mu_i(t) N_i)/T(t)]
\end{equation}
contains the Lagrange multipliers $\mu_i$, $T$ which are in general time dependent. 
They are not identical to the equilibrium parameters temperature and chemical potentials, but can be regarded as their generalizations in non-equilibrium. 

If the relaxation time $\tau_{\rm r}$ of any fluctuation in the state described by $\rho_{\rm rel}(t)$ is short, 
the limit $\epsilon \to 0$ in Eq. (\ref{rhoZ}) can be replaced by $\epsilon \le 1/\tau_{\rm r}$, the memory of the system is short (Markov limit). 
However, fluctuations $B_l$ for which $\dot \lambda_l/\lambda_l \le 1/\tau_{\rm r}$ should be considered as new relevant observables, the relevant distribution based on the former set $\{ B_i \}$ of relevant observables freezes out. 
The temporary evolution of the system in Markov approximation requires a larger set of relevant observables which includes the slow observables $B_l$.
For instance, the occupation numbers of single-particle states may be included to obtain reaction-kinetic equations for the distribution function describing the evolution of the system after (chemical) freeze-out of the thermodynamic equilibrium.

\section{Blast-wave model of particle freeze-out}
\label{sec:blast_wave}

To describe the system formed in a heavy-ion collision, one must consider the fast system expansion. However, instead of introducing the collective velocity as another relevant observable and deriving the hydrodynamic equations from the NSO \cite{Harutyunyan:2021rmb}, we employ the blast-wave model parametrization in its standard form \cite{BWModel}, assuming particles are emitted from the boost-invariant and cylindrically symmetric hypersurface $\Sigma_\mu$ with the volume element $d\Sigma_\mu$ given by:
\begin{equation}
    d\Sigma_\mu = \left(\tau r \cosh\eta, 0, 0, -\tau r \sinh\eta\right) dr d\phi d\eta,
\end{equation}
at a constant value of the proper time $\tau = \sqrt{t^2-z^2} = {\rm const}$. Here $r$ is the transverse radius, $\phi$ is the azimuthal angle in coordinate space, and $\eta = \tanh^{-1}(z/t)$ is the space-time rapidity. In this model, $r\in\left[0; R\right]$ and $\eta\in\left[-\eta_{max};\eta_{max}\right]$, with $R$ and $\eta_{max}$ being the system size in transverse and longitudinal directions, respectively.
The collective flow velocity in midrapidity is parametrized as
\begin{equation}
    u^{\mu}(r, \eta=0) = \left(\cosh \rho, \sinh\rho \cos \phi, \sinh\rho \sin \phi, 0\right),
    \label{eq:velocity}
\end{equation}
where $\rho$ is the velocity profile:
\begin{equation}
	\rho = \tanh^{-1} \left[ \beta_S \left(\dfrac{r}{R}\right)^n\right],
\end{equation}
with the profile exponent $n$ and transverse expansion velocity at the surface $\beta_S$. The full velocity field is obtained by boosting \eqref{eq:velocity} in the $\eta$ direction. Having the freeze-out hypersurface and collective velocity fields defined, the particle invariant momentum distribution is given by the Cooper-Frye formula \cite{Cooper_Frye}:
\begin{equation}
	E\dfrac{d^3N_i}{d^3p} = \dfrac{g_i}{(2\pi\hbar)^3} \int d\Sigma_\mu p^\mu f_i(p^\mu u_\mu),
    \label{eq:CF_formula}
\end{equation}
where $g_i$ is the spin degeneracy factor of the $i$-th particle species, $p_\mu$ is the particle four-momentum, and $f_i(E)$ is the particle distribution function:
\begin{equation}
	f_i(E) = \left( \exp\left[ \dfrac{E - \mu_i}{T} \right] + a_i \right)^{-1},
\end{equation}
with $a_i = +1$ for fermions and $a_i = -1$ for bosons, and $\mu_i$ and $T$ being the particle chemical potential and freeze-out temperature or their nonequilibrium generalizations from the NSO method if nonequilibrium is assumed.
Then the resulting particle distribution function is given as:
\vspace{-2mm}
\begin{widetext}
    \begin{equation}
		\dfrac{d^6N_i}{dp_Tdyd\eta dr d\phi d\psi} = \dfrac{g_i \tau r p_T m_T}{(2\pi\hbar)^3} \cosh \left(y - \eta\right) \left( \exp \left[\frac{m_T \cosh \rho \cosh(y - \eta) - p_T \sinh \rho \cos(\phi - \psi) - \mu_i}{T}\right]  + a_i \right)^{-1},
        \label{eq:dist_func}
	\end{equation}
\end{widetext}
where $p_T$ the transverse momentum of the particle, $m_T$ the transverse mass of the particle, $\psi$ is the azimuthal angle in momentum space and $y$ is the rapidity of the particle.

For resonances, we include additional mass attenuation in the form of the nonrelativistic Breit-Wigner distribution with constant width. Since particle-antiparticle ratios in the energy range of interest are approximately unity over a broad transverse momentum range, we set baryon, strangeness, and electric charge chemical potentials to zero. This reduces the number of free parameters while maintaining a reasonable fit quality. Then the particle chemical potential is defined as
\begin{equation}
    \mu_i = \mu_\pi \delta_{i,\pi} + \mu_K \delta_{i,K},
\end{equation}
where $\mu_\pi$ and $\mu_K$ are pion and kaon nonequilibrium effective chemical potentials.

\begin{table}[htb]
    \begin{tabular}{|c|W{c}{5mm}|W{c}{5mm}|W{c}{5mm}|W{c}{5mm}|W{c}{5mm}|c|c|c|c|}
		\hline
		\multirow{4}{*}{\rotatebox{90}{Model}} & \multicolumn{5}{c|}{\multirow{3}{*}{\backslashbox[33mm]{Lagrange\\ multipliers}{Approximations}}} & \multicolumn{3}{c|}{Kinetic} & Final \\
		 & \multicolumn{5}{c|}{} & (a) & (b) & (c) & decays \\
		 & \multicolumn{5}{c|}{} & ideal & res. & in-med. & + \\
		 & $T$ & $\mu_p$ & $\mu_n$ & $\mu_\pi$ & $\mu_K$ & $p,\pi,K$ & gas & effects & scatterings \\
		\hline
		A & \checkmark & \checkmark & \checkmark & - & - & - & - & - & - \\
		B & \checkmark & \checkmark & \checkmark & \checkmark & \checkmark & - & - & - & - \\
		C & \checkmark & \checkmark & \checkmark & - & - & \checkmark & \checkmark & - & \checkmark \\
		D & \checkmark & \checkmark & \checkmark & \checkmark & - & \checkmark & \checkmark & - & \checkmark \\
		\hline
	\end{tabular}
    \caption{Approximations corresponding to the models A-D for describing particle production in HIC, (a)-(c) concern the treatment of the spectral function (\ref{selfc}), for details see text.}
    \label{tab:approx}
\end{table}

\section{Results and discussion}\label{sec:results}

To solve the puzzle of the low-momentum enhancement of the pion spectra in HIC at ultrarelativistic energies \cite{ALICE:2020nkc}
we apply the method of the NSO. 
We can include the numbers $N_\pi$ and $N_K$ of pions/kaons into the set of relevant variables since they are long-living fluctuations, with lifetimes determined by weak interaction.

\begin{figure}[!ht]
\includegraphics[width=0.95\columnwidth]{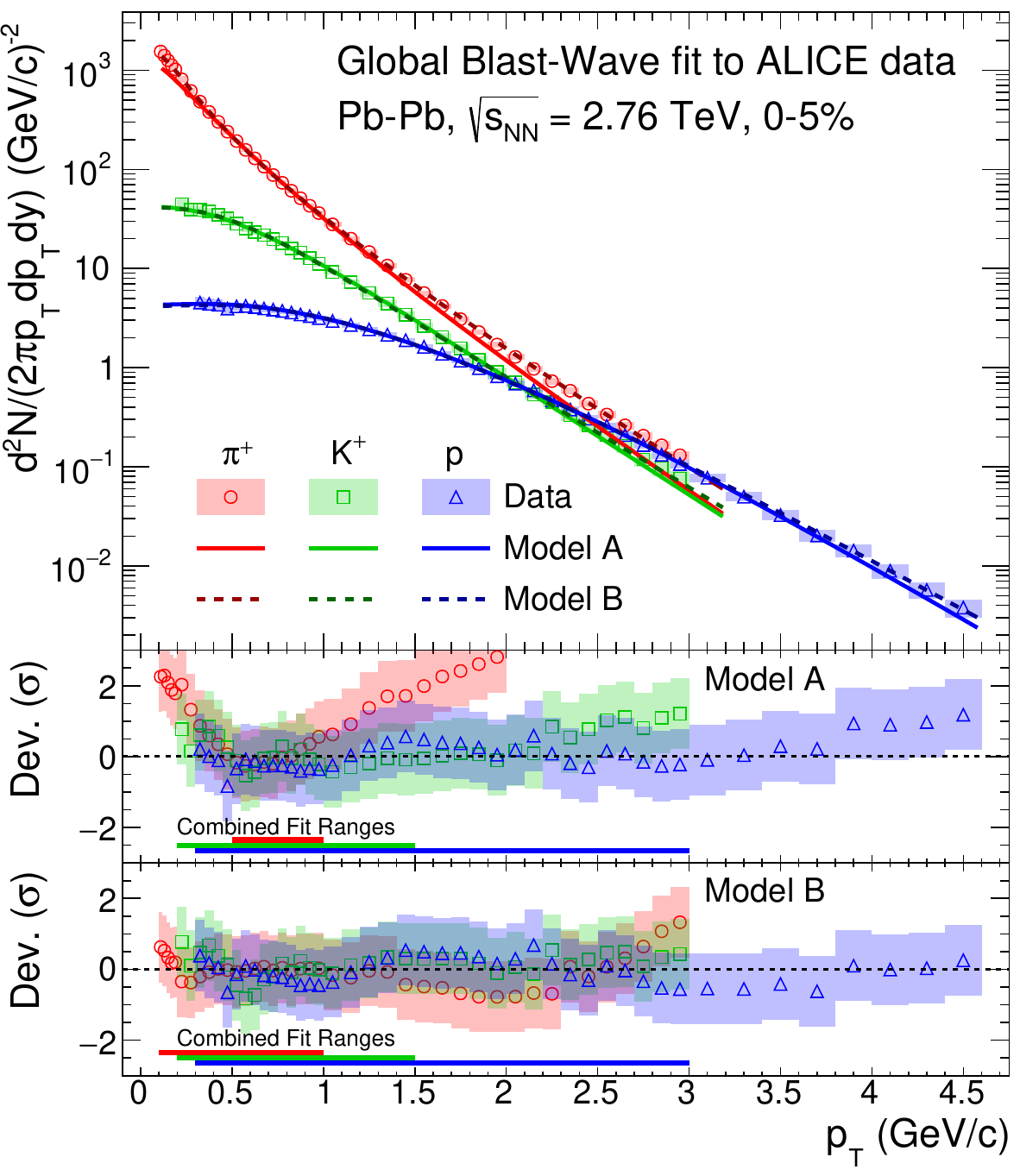}
\caption{Transverse-momentum spectra of $\pi^{+}$, $K^{+}$, and $p$ (anti-particles are not shown) measured by the ALICE collaboration~\cite{ALICE-2013} in 0-5\% central PbPb collisions at $\sqrt{s_{NN}} = 2.76$~TeV compared to the  
{standard} blast-wave fit and to the fit with nonequilibrium distribution function.}
\label{fig:simple_fit}
\end{figure}

Different approximations are considered (see Table \ref{tab:approx}): 
Model A is the standard blast-wave model without hadron resonances and their decays, no final state rescattering applied for the description of pion, kaon and proton spectra at the kinetic freezeout, see Fig. \ref{fig:simple_fit}. A strong deviation from ALICE data at low momenta is observed. 
The conventional approximation A gives similar results to those reported in the ALICE publication \cite{ALICE-2013}. Namely, the temperature is about 95 MeV, and the surface velocity is about 87\% of the speed of light.

In model B, we introduce the non-equilibrium pion and kaon chemical potentials $\mu_\pi$ and $\mu_K$ into the Bose distribution \cite{Kataja:1990tp,Blaschke:2020afk} which enters the blast-wave function \eqref{eq:dist_func}. 
The fit result is shown in Fig. \ref{fig:simple_fit} and the fit parameters are summarized in Table \ref{tab:simple_fit}. 
Compared to model A, the fit gives a significantly higher temperatures, i.e. about 120 MeV at roughly the same flow velocity.
The pion and kaon chemical potentials are high and close to the particle masses with $\mu_\pi = (126.2 \pm 3.8)$ MeV and $\mu_K = (402.9 \pm 22.6)$ MeV.
One can see that such an extension gives not only the correct pion/kaon particle density but also the spectra using just two additional parameters. 

The blast-wave function \eqref{eq:dist_func} is fitted to the data by minimizing the $\chi^2$ over the combined fit range. Following the experimental procedure~\cite{ALICE-2013,Braun-Munzinger:2018hat,ALICE:2020nkc}, for model A we set the combined fit $p_T$ ranges to $0.5-1.0$ GeV/c for $\pi^{\pm}$, $0.2-1.5$ GeV/c for $K^{\pm}$, and $0.3-3.0$ GeV/c for\\ (anti-)protons \cite{ALICE-2013}. 
{For model B to be sensitive to changes in $\mu_\pi$, we extend the $p_T$ fit range for $\pi^{\pm}$ to $0.1-1.0$ GeV/c. }
To examine the overall fit quality, we introduce $\chi^{2*}$, which is the $\chi^2$ function evaluated for all spectra over the $p_T \in [0,2]$ GeV/c.
It is remarkable that the fit with model B describes the data over a broad $p_T$ range, even outside of the fit ranges since $\chi^{2*}$ is $6.7$ times better in the extended model with finite effective chemical potentials than in model A.
The NSO-based approximation B with non-equilibrium pion and kaon chemical potentials provides an excellent description of the particle spectra including the low-momentum enhancement.

\begin{figure*}[!ht]
  \includegraphics[width=0.96\textwidth]{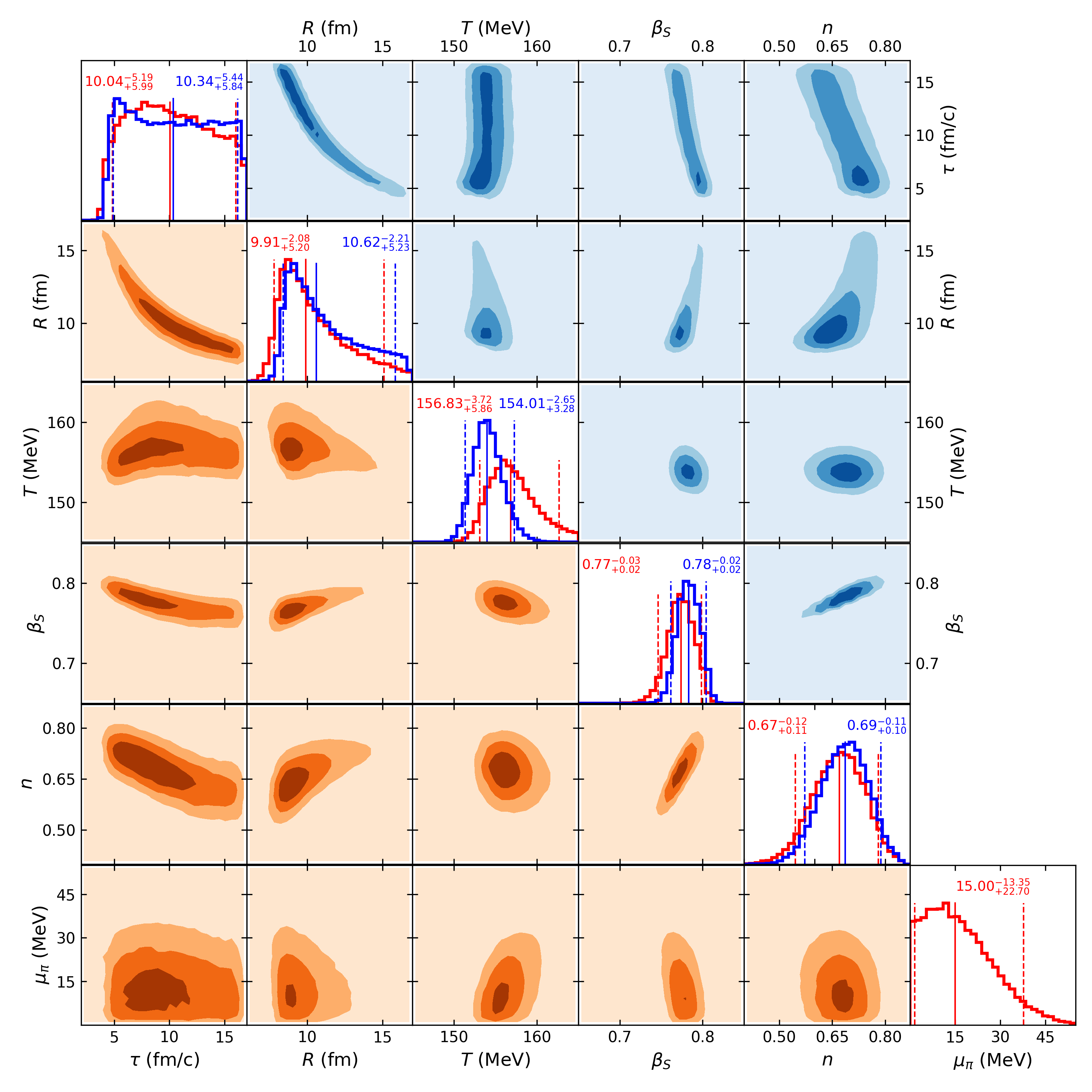}
  \caption{Marginal (on diagonal) and joint marginal posterior probability distributions of model parameters for {model C (blue, upper triangle) and model D (orange, lower triangle) of the overpopulated pion states.} 
Solid and dashed vertical lines and numeric values indicate median values and 90\% credible intervals of the parameter distributions, respectively.}
  \label{fig:square_plot}
\end{figure*}

\begin{table*}[ht]
    \centering
    \begin{tabular}{lccccccccc}
        \toprule 
        Model & $\tau$ (fm/c) & $R$ (fm) & $T$ (MeV) & $\beta_{S}$ & $n$ & $\mu_\pi$ (MeV) & $\mu_K$ (MeV) & $\chi^2$/ndof & $\chi^{2*}$/ndof \\
        \midrule 
        A) KFO & - & - & $97.7 \pm 3.9$  & $0.879 \pm 0.005$ & $0.709 \pm 0.019$ & - & - & $0.148$ & $0.956$ \\
        B) KFO & - & - & $118.1 \pm 3.7$ & $0.858 \pm 0.005$ & $0.662 \pm 0.023$ & $126.2 \pm 3.8$ & $402.9 \pm 22.6$ & $0.136$ & $0.143$ \\
        C) CFO & $8.88$ & $11.54$ & $154.02$ & $0.786$ & $0.699$ & - & - & $0.1$ & $0.1$ \\
        D) CFO & $8.08$ & $11.5$ & $155.98$ & $0.783$ & $0.697$ & $9.52$ & - & $0.1$ & $0.1$ \\
        \bottomrule 
    \end{tabular}
    \caption{Results of the combined blast-wave fits of identified particle spectra at kinetic freeze-out (KFO) and the MAP estimate of the model parameters from the Bayesian analysis at the chemical freeze-out (CFO) with and without pion/kaon chemical potential. $\tau, R, \beta_S, n$ are parameters of the blast-wave model, Eq. (\ref{eq:dist_func}).}
    \label{tab:simple_fit}
\end{table*}

\begin{figure}[h]
\includegraphics[width=0.95\columnwidth]{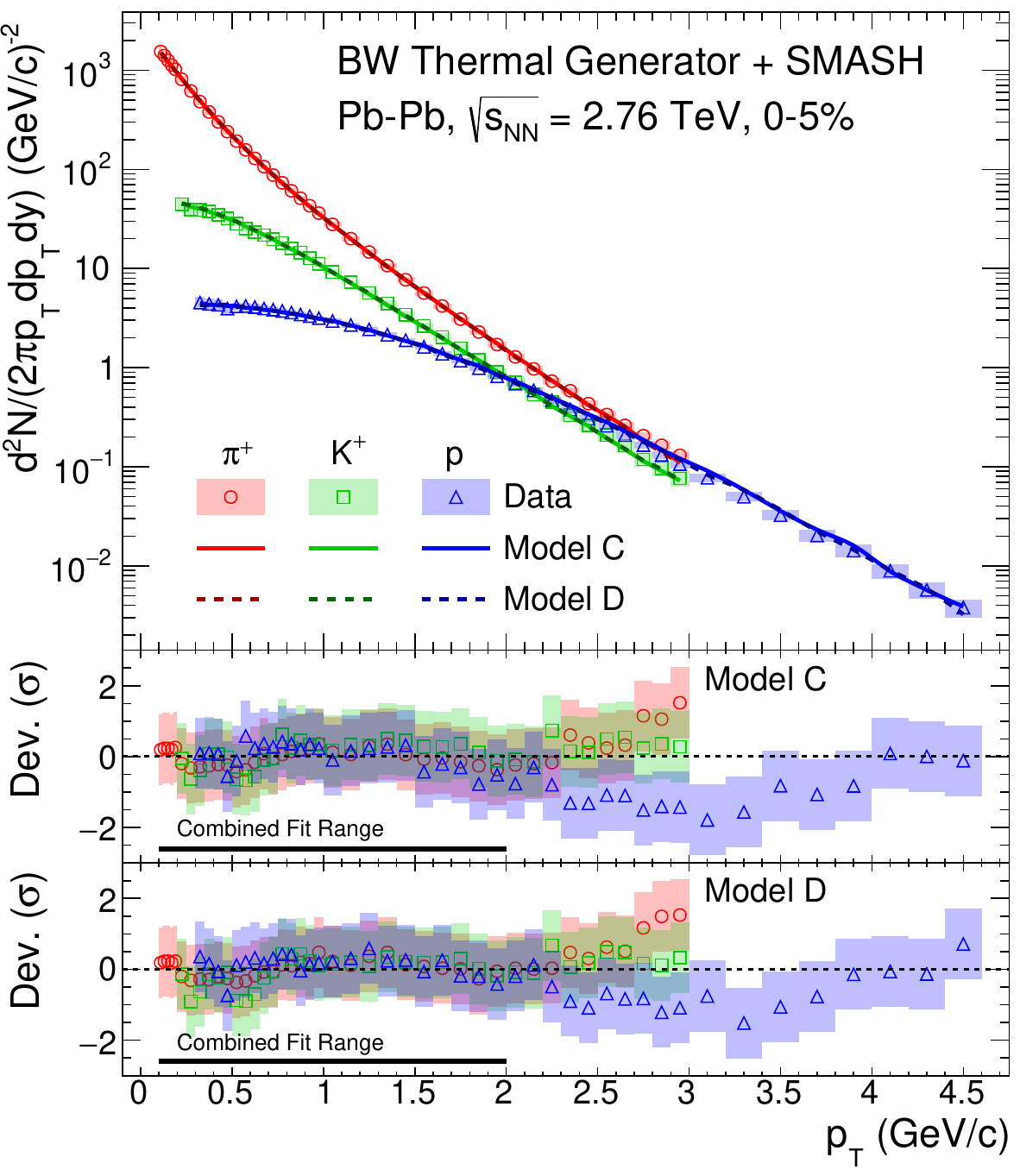}
\caption{Transverse momentum spectra of pions, kaons, and protons (anti-particles are not shown) compared to the MAP estimate from models C and D.}
\label{fig:map_smash_mu}
\end{figure}

In the approximations C and D of Table \ref{tab:approx}, we explicitly account for resonance decays and final state interactions between hadrons. This is an attempt to extend the description of the system to a stage before the kinetic freeze-out, namely to the chemical freeze-out stage. For this purpose, we developed a blast-wave-based thermal particle generator coupled with the SMASH hadronic transport model \cite{smash}. First, for a given set of model parameters, we sample the particles' positions, momenta, and, for unstable states, masses on the freeze-out hypersurface using the previously obtained distribution function \eqref{eq:dist_func}. We use all hadronic states from the SMASH particle list that do not contain charm and bottom quarks. Then, the particles generated from the freeze-out hypersurface are input into the SMASH model which serves as an afterburner, performing resonance decays, elastic scatterings, and other final state interactions.
Taking into account the nonzero value of the effective pion chemical potential at kinetic freeze-out, we challenge the assumption of the chemical equilibrium on the hadronization surface and consider two cases: model C with full chemical equilibrium and model D with the overpopulated pion state,  modeled by the inclusion of the effective pion chemical potential.

To infer the model parameters we use the Bayesian analysis workflow similar to one used in \cite{Jussi_Bayesian}. 
Since the full model is computationally expensive, we replace it with Gaussian process emulators and reduce the number of emulators needed using the principal component analysis. 
Uniform priors and multivariate normal likelihood were used in this analysis. The marginal posterior probability distributions of the model parameters were obtained using the Markov chain Monte Carlo sampling. They are shown in Fig. \ref{fig:square_plot} together with the median values and 90\% credible intervals, see also table \ref{tab:simple_fit} for maximum a posteriori probability (MAP) estimate of the parameter values. 
The transverse momentum spectra for models C and D are shown in Fig. \ref{fig:map_smash_mu}.
The Bayesian analysis using a blast-wave model coupled to SMASH gives a posterior temperature that is about 156 MeV, which is consistent with lattice QCD \cite{Borsanyi2020,Bazavov2019} and statistical hadronization models \cite{ALICE2018_He,Andronic:2017pug,ALICE:2022wpn}, 
and a mean flow velocity of about 77\% of the speed of light.
Repeating the same analysis with model D results in the same fit quality and a value for $\mu_\pi$ that is compatible with zero. This is confirmed by the fact that corresponding Bayes factor \cite{Everett2021}  $B^{D}_{C} = 0.434$ is of the order unity, while the posterior distributions for model D are getting broader which is seen in Fig. \ref{fig:square_plot}.

We show that both models B and C describe the situation of the overpopulation of the low-momentum pion states, but in different approximations. 
After the chemical freeze-out, inelastic collisions between hadrons no longer contribute to the changes in occupation numbers, while resonances still can decay and (quasi-)elastic processes lead to thermal equilibration and modify hadron spectra until kinetic freeze-out. 
During this stage, the {effective} number of pions/kaons remains nearly conserved {in the sense that a hadronic resonance that decays into a number of pions may be regarded as a reservoir for that same number of additional pions} \cite{Bebie1992}. Since pion/kaon decay is a weak process which is sufficiently slow, the pion number $N_{\pi}$ and kaon number $N_K$ must be included in the set of relevant observables $\{B_i\}$. This leads to the appearance of the effective chemical potential $\mu_i$ in the distribution function \eqref{eq:dist_func} as a consequence of the self-consistency condition \eqref{selfc}. An explicit treatment of resonance decays and final state interactions using SMASH confirms this result. 

\section{Conclusions}\label{sec:conclusions}

In conclusion, models B, C, and D provide reasonable approximations for describing the dynamics of the expanding fireball. 
The model C offers a microscopic description of the collision processes and resonance decays including rescattering, albeit with free-space parameters without medium effects,
starting at the chemical freeze-out surface and continuing through to kinetic freeze-out. 
The extension of the relevant distribution introducing the nonequilibrium Lagrange parameter $\mu_\pi$ in model D gives no essential improvement because the mesonic overpopulation at low-$p_T$ is already described by the decay of resonances and their thermalization. 
In contrast, model B reflects this overpopulation of low-$p_T$ meson states after thermalization, without following microscopically the thermalization process.
The results show that this simple extension of the blast-wave model with mesonic effective chemical potentials presents an efficient alternative for describing particle spectra at kinetic freeze-out. 
The simple model B can also be improved considering a better approximation of the spectral function evaluating Eq. (\ref{selfc}) with the extended set of Lagrange parameters, taking correlations into account.

However, all models, in particular C and D, are also approximations since they do not consider in-medium effects, for instance the Mott effect \cite{Blaschke:2024jqd}, which are essential at high densities. 
To be consistent with the current calculations explaining the particle yields obtained from high-precision experiments \cite{Andronic:2018qqt,Donigus:2022haq}, which explicitly show the role of in-medium corrections, an improved treatment of spectra with in-medium corrections is also required, left for future work.

In a next step of the development of Zubarev's NSO approach to describe the nonequilibrium evolution of the hadron resonance gas created in heavy-ion collisions would be the extension of the set of observables \eqref{selfc} to include the flow of particle species and to describe the formation of resonances as a result of interactions to be included to the model Hamiltonian.
The goal of such an extended NSO approach would be to provide a consistent scenario for hadron production in heavy-ion collisions which shall justify the success of the thermal statistical model and its limitations.
To give an example, the fugacity parameters that are introduced in thermal statistical models on an ad hoc basis to improve the description of experimental data (see, e.g. \cite{Begun_PRC_2014,Petran:2013lja}) could be related to the Lagrangian multipliers in the NSO formalism and would thus appear as the result of a consistent and systematic approach applied to describe the nonequilibrium system.

\begin{acknowledgments}
We acknowledge encouraging discussions with Peter Braun-Munzinger and Krzysztof Redlich. 
The authors are grateful to Pasi Huovinen for his valuable comments.
D.B. was supported by NCN under grant No. 2021/43/P/ST2/03319, O.V. received support from NCN under grant No. 2022/45/N/ST2/02391 and G.R. acknowledges a stipend from the Foundation for Polish Science within the Alexander von Humboldt programme under grant No. DPN/JJL/402-4773/2022. B.D. acknowledges support from Bundesministerium f\"{u}r Bildung und Forschung through ErUM-FSP T01, F\"{o}rderkennzeichen 05P21RFCA1. Calculations have been carried out using resources provided by Wroclaw Centre for Networking and Supercomputing, grant No. 570.
\end{acknowledgments}

\bibliography{document}

\end{document}